\documentclass[prd,aps,footinbib,longbibliography,twocolumn]{revtex4-2}
\usepackage[babel]{csquotes}
\usepackage{amsmath,amssymb}
\usepackage[colorlinks,citecolor=blue,linkcolor=blue,urlcolor=blue]{hyperref}
\usepackage{graphicx}
\usepackage{dcolumn}
\usepackage{bm}
\usepackage{physics} 
\usepackage{bbold}
\usepackage{mathtools}
\usepackage{xcolor}
\usepackage{tikz}
\usepackage{tikz-cd}
\usepackage{comment}
\usepackage{subfigure}
\usepackage{soul}
\usepackage{framed}
\usepackage{mdframed}
\usepackage{appendix}
\usepackage{hyperref}
\usepackage{natbib}
\usepackage[normalem]{ulem}
\usepackage[T1]{fontenc}
\usepackage{float}
\usepackage{amsthm}

\def\bra#1{\mathinner{\langle{#1}|}}
\def\ket#1{\mathinner{|{#1}\rangle}}

\def\be{\begin{equation}}
\def\ee{\end{equation}}

\def\ea{{\mathcal{E}_A}}
\def\eb{{\mathcal{E}_B}}

\newcommand{\M}{\mathcal{M}}

\newtheorem{definition}{Definition}

\usepackage{xcolor}
\usepackage{xspace}
\usepackage{ifthen}

\newcommand{\showcomments}{false}

\newcommand{\marios}[1]%
{\ifthenelse{\equal{\showcomments}{true}}%
{{\color{blue}{\small \textbf{M:} #1}}}{\xspace}}%

\newcommand{\caslav}[1]%
{\ifthenelse{\equal{\showcomments}{true}}%
{{\color{red}{\small \textbf{C:} #1}}}{\xspace}}%

\newcommand{\acd}[1]%
{{\color{magenta}{#1}}}{\xspace}%

\newcommand{\viktoria}[1]%
{{\color{orange}{#1}}}{\xspace}%

\newcommand{\new}[1]
{{\color{teal}{#1}}}{\xspace}%

\newcommand{\revise}[1]%
{{\color{blue}{#1}}}{\xspace}%

\begin{document}
\title{Indefinite causal order and quantum coordinates}

\author{Anne-Catherine de la Hamette}
\affiliation{University of Vienna, Faculty of Physics, Vienna Doctoral School in Physics, and Vienna Center for Quantum Science and Technology (VCQ), Boltzmanngasse 5, A-1090 Vienna, Austria}
\affiliation{Institute for Quantum Optics and Quantum Information (IQOQI),
Austrian Academy of Sciences, Boltzmanngasse 3, A-1090 Vienna, Austria}

\author{Viktoria Kabel}
\affiliation{University of Vienna, Faculty of Physics, Vienna Doctoral School in Physics, and Vienna Center for Quantum Science and Technology (VCQ), Boltzmanngasse 5, A-1090 Vienna, Austria}
\affiliation{Institute for Quantum Optics and Quantum Information (IQOQI),
Austrian Academy of Sciences, Boltzmanngasse 3, A-1090 Vienna, Austria}

\author{Marios Christodoulou}
\affiliation{University of Vienna, Faculty of Physics, Vienna Doctoral School in Physics, and Vienna Center for Quantum Science and Technology (VCQ), Boltzmanngasse 5, A-1090 Vienna, Austria}
\affiliation{Institute for Quantum Optics and Quantum Information (IQOQI),
Austrian Academy of Sciences, Boltzmanngasse 3, A-1090 Vienna, Austria}

\author{\v{C}aslav Brukner}
\affiliation{University of Vienna, Faculty of Physics, Vienna Doctoral School in Physics, and Vienna Center for Quantum Science and Technology (VCQ), Boltzmanngasse 5, A-1090 Vienna, Austria}
\affiliation{Institute for Quantum Optics and Quantum Information (IQOQI),
Austrian Academy of Sciences, Boltzmanngasse 3, A-1090 Vienna, Austria}

\date{\small October 1, 2025}

\begin{abstract}
Classically the causal order of two timelike separated events A and B is fixed -- either A before B or B before A. This is no longer true in quantum theory, where it is possible to encounter superpositions of causal orders. The quantum switch is one of the most prominent processes with indefinite causal order. Optical realizations of the quantum switch have been successfully implemented in experiments, but, some argue this merely simulates a process with indefinite causal order and that a superposition of spacetime metrics is required for a true realization. Here, we provide a relativistic definition of causal order between operationally defined events that defines a meaningful observable in both the general relativistic and quantum mechanical sense.
Importantly, this observable does not distinguish between the indefinite causal order implemented on an optical bench and the gravitational quantum switch, a gedankenexperiment where the indefinite causal order is achieved by a quantum superposition of gravitational fields. Therefore, our results support the thesis that the optical quantum switch is just as much a realization of indefinite causal order as its gravitational counterpart, which makes use of the quantum mechanical behavior of spacetime.
\end{abstract}

\maketitle

The study of indefinite causal order (ICO) has seen remarkable progress over the last years both theoretically \cite{Hardy_2005,Chiribella_2013, Oreshkov_2012, zych_2019} and experimentally \cite{Procopio_2015, Rubino_2017, Rubino_2022, Goswami_2018, Wei_2019, Taddei_2021}. The field was inspired by the goal to combine general relativity and quantum theory \cite{Hardy_2005} as well as the provable advantage of ICO over causal quantum circuits in quantum information processing \cite{Chiribella_2013, Guerin_2018, kristjansson2024}. General relativity is a deterministic theory with a dynamical causal structure. Quantum theory is a probabilistic theory with a fixed causal structure. Thus, combining the two would allow for processes to which no definite causal ordering can be assigned. 

One of the simplest and most prominent information theoretic primitives of ICO is the quantum switch \cite{Chiribella_2013, Oreshkov_2012}. Depending on the state of a control qubit, two operations $U_A$ and $U_B$ are applied to a target system in a superposition of `$U_A$ before $U_B$' and `$U_B$ before $U_A$'.
Concretely, denoting the initial state of the target system by $\ket{\psi}_T$ and taking the operations to be unitary, the resulting state is
\begin{equation}
\frac{1}{\sqrt{2}}(\ket{0}_C U_BU_A\ket{\psi}_T+\ket{1}_C U_AU_B\ket{\psi}_T),
\end{equation}
where $\ket{0}_C$ and $\ket{1}_C$ are two orthogonal states of the control system. Because this evolution cannot be described by a sequence of both operations in a definite causal order or a classical mixture thereof, we say that the causal order of `events' $A$ and $B$ is indefinite \cite{Oreshkov_2016_Cerf}.

Various ways of implementing ICO have been reported and described in the literature. In the \emph{optical quantum switch}, the target system is passing through the setup in a superposition of paths \cite{Procopio_2015, Rubino_2017, Rubino_2022}. This type of ICO  has been successfully implemented on the optical bench \cite{Goswami_2018, Wei_2019, Taddei_2021}. A different implementation is provided by the \emph{gravitational quantum switch} \cite{zych_2019, Paunkovic_2020}. In this gedankenexperiment, a massive object in superposition of two locations sources a gravitational field in superposition, while two agents perform operations on a test particle. A minimal assumption of most quantum gravity theories is that the spacetime metric sourced by a mass in superposition is itself in superposition.  This regime, where matter behaves quantum mechanically and spacetime is in a quantum superposition of semi-classical states, has been studied extensively as a possible interface between quantum information theory and quantum gravity (e.g.~\cite{Marletto_2017, Bose_2017, Belenchia_2018, Christodoulou_2019, Anastopoulos_2020, zych_2019, delahamette2021falling, aspelmeyer2021}). Then, choosing an appropriate setup can result in an indefinite causal order due to the different time dilations experienced in each branch of the superposition. 

The radically different physics involved in the optical and the gravitational implementations of the quantum switch leads some to question whether it can be claimed that `genuine' or `proper' ICO has been experimentally realized. The quantum information community is roughly divided along two theses \cite{Procopio_2015, Oreshkov_2019, Paunkovic_2020, Felce_2022, Vilasini_2022, Ormrod2023causal}. On the one hand, those that claim that the optical switch \enquote{can be seen as the first \emph{realization} of a superposition of causal orders} \cite{Procopio_2015},  and on the other, those who view it as a mere \emph{simulation} of ICO. The latter camp argues that \enquote{the current quantum switch experimental implementations do not feature superpositions of causal orders between spacetime events}, and that \enquote{these superpositions can only occur in the context of superposed gravitational fields} \cite{Paunkovic_2020}. This debate is closely intertwined with the question whether realizing `genuine' ICO might be employed in the future to witness non-classicality of spacetime.\\

To bring clarity into this debate, we cast the notion of ICO in a general relativistic language. This allows us to study the indefiniteness of causal order in a unified setting for both variations of the quantum switch. We first discuss the definition of an event in the context of a superposition of spacetimes and propose an operational, diffeomorphism-invariant notion in terms of worldline coincidences. We then identify a single binary quantity that encodes the causal order of a process, the sign of the proper time, and demonstrate that it is invariant under \emph{quantum-controlled diffeomorphisms} --- a quantum-controlled arbitrary and independent change of coordinates in each branch of the superposition. We further show that this quantity is not only an operationally meaningful observable in the general relativistic but also in the quantum mechanical sense. Crucially,  this observable does not differentiate whether ICO is arising due to a superposition of paths on a definite metric or a superposition of spacetimes, that is, it \emph{cannot} distinguish between the optical and the gravitational quantum switch. 
This strengthens the claim that the optical and gravitational implementation of the quantum switch exhibit the same `type' of ICO. 

This work is motivated by recent research on quantum reference frames \cite{hardy2019, Giacomini_2019, loveridge_symmetry_2018, vanrietvelde2018change, castroruiz2019time, hoehn2019trinity, delahamette2020, Krumm_2021, castroruiz2021relative, delahamette2021perspectiveneutral, delahamette2021entanglementasymmetry, Cepollaro_2021, delahamette2021falling, Kabel2022conformal}, which suggests that the invariance of physical laws under coordinate transformations extends to quantum superpositions thereof. Indefinite causal order, as we define it here, is a prototype case of a physically meaningful quantity that has no classical counterpart and is an invariant under quantum coordinate changes.

\emph{Definite causal order ---} Causal order is defined with respect to signalling conditions in quantum information theory. We suggest a simple definition of definite causal order based on proper time, inspired by and applicable to the several setups considered in the study of ICO \cite{Chiribella_2013, Oreshkov_2012, Procopio_2015, Rubino_2017, Rubino_2022, Goswami_2018, Wei_2019, Taddei_2021, zych_2019, castroruiz2019time}.

\begin{figure}
    \centering
    \includegraphics[width=0.25\textwidth]{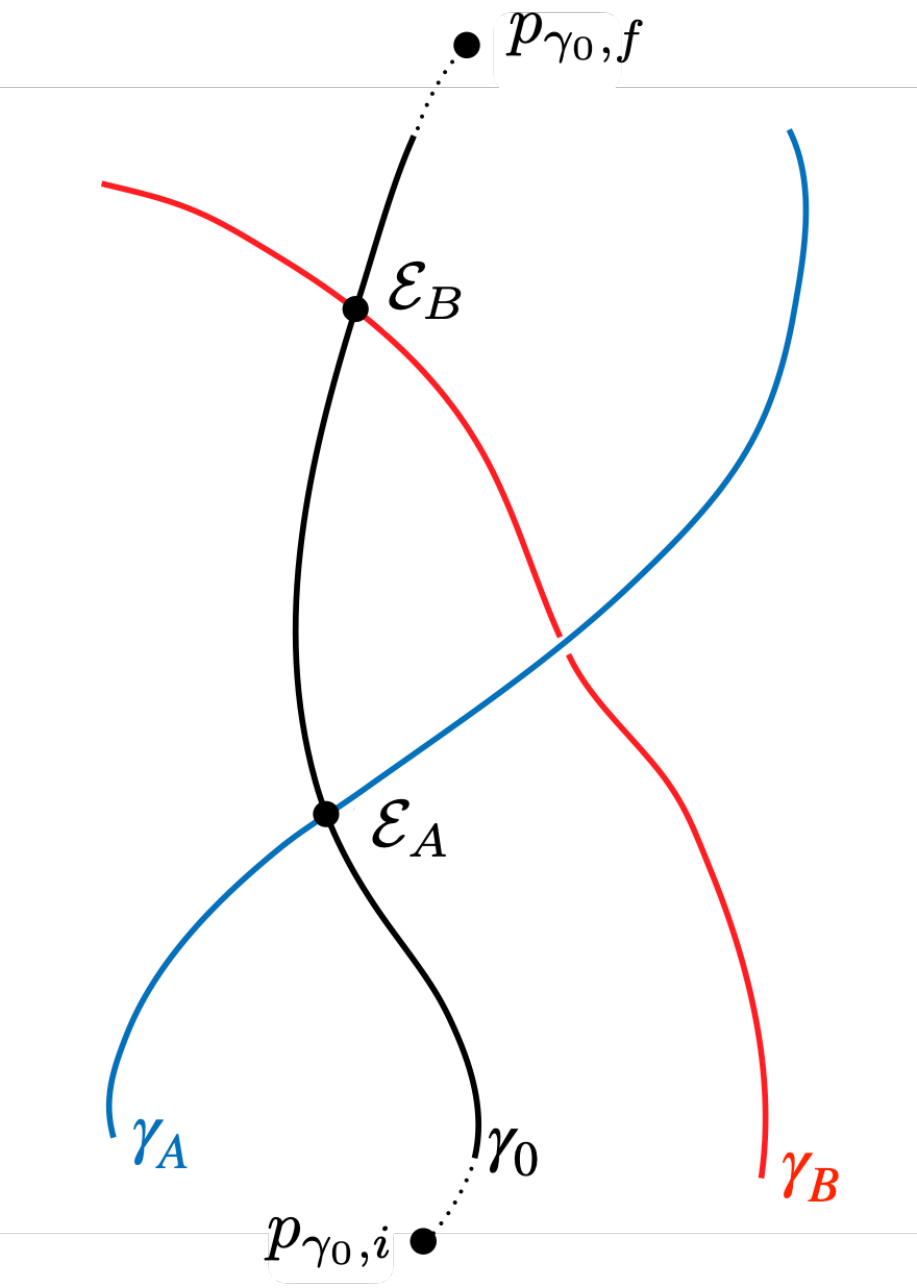}
    \caption{We consider a scenario of two systems and a test particle on a fixed spacetime. The timelike worldline $\gamma_0$ of the test particle is depicted in black, while the timelike curves of the systems are depicted by the blue line $\gamma_A$ and the red line $\gamma_B$. The initial and final points defining the curve $\gamma_0$ are denoted by $p_{\gamma_0,i}$ and $p_{\gamma_0,f}$, respectively. The worldlines $\gamma_0$ and $\gamma_A$ coincide once and their crossing defines the event $\ea$, while $\eb$ is defined by the single crossing of the worldlines $\gamma_0$ and $\gamma_B$. We can use the proper time of $\gamma_0$ together with fixed initial and final points to define a causal order for the events.}
    \label{fig:worldlines}
\end{figure}

Take two systems and a test particle interacting with each of the systems once. We take each interaction to define an \emph{event}. The events may correspond to the application of a unitary operation acting on the electron spin. We model this as three timelike curves $\gamma_0$, $\gamma_A$, and $\gamma_B$, with $\gamma_0$ coinciding once with $\gamma_A$ and once with $\gamma_B$ at the coincidences $\mathcal{E}_A$ and $\mathcal{E}_B$ (see Fig.~\ref{fig:worldlines}). Coincidences of worldlines define events in a coordinate independent way. Note that while, classically, each such event  is associated with a single spacetime point -- sometimes referred to as an event in other works -- this is will no longer be the case when considering superpositions of spacetimes. Therefore, we reserve the term `event' for events defined in terms of worldline coincidences and refer to spacetime points or manifold points otherwise.

Each curve can be parametrized by its proper time,
\begin{align}
    \tau_{\gamma_0} = \frac{1}{c}\int_{p_{\gamma_0,i}}^{p_{\gamma_0,f}}\sqrt{-g_{\mu\nu}dx^\mu dx^\nu},\label{eq:properTime}
\end{align}
elapsed between two points $p_{\gamma_0,i}$ and $p_{\gamma_0,f}$. We take $p_{\gamma_0,i}$ as the event at which the test particle is released and $p_{\gamma_0,f}$ as the one at which a final measurement is applied. This  fixes a time orientation along $\gamma_0$. $\eb$ lies either in the future or past lightcone of $\ea$. With our time orientation, $\eb$ lies in the future of $\ea$ if $\tau_{A}<\tau_{B}$. This corresponds to being able to signal from $\ea$ to $\eb$, which is also the operational notion of causal order employed in quantum information. Therefore, the causal order is encoded in the sign of the proper time difference.   Define the crossing points $p_{\mathcal{E}_A} \equiv \gamma_0(\tau_{A})$, $p_{\mathcal{E}_B} \equiv \gamma_0(\tau_{B})$. The distinction between the events $\mathcal{E}_X$ and points $p_{\mathcal{E}_X}$ will be important in what follows.

\begin{definition}[Causal order between events]\label{def:CO}
    Consider two events $\mathcal{E}_X$, $X\in\{A,B\}$, connected by a timelike worldline $\gamma_0$ with crossing points $p_{\mathcal{E}_X} \equiv \gamma_0(\tau_{X})$. The causal order between $\ea$ and $\eb$ is the sign $s\equiv\mathrm{sign}(\Delta \tau)$ of the proper time difference $\Delta \tau \equiv \tau_{B}-\tau_{A}$.
\end{definition}
Note that this definition assumes that the events are uniquely defined, i.e.~that the worldline $\gamma_0$ crosses both $\gamma_A$ and $\gamma_B$ exactly once.

\emph{Events on a superposition of spacetimes --- } Consider two globally hyperbolic spacetimes $g^{(1)}$ and $g^{(2)}$ each associated to a manifold $\mathcal{M}_i$, $i\in\{1,2\}$, diffeomorphic as differentiable manifolds. We take the same setup as above on each spacetime, with two events $\mathcal{E}_A^{(i)}$, $\mathcal{E}_B^{(i)}$ defined as coincidences. Next, we consider a superposition of two semi-classical, macroscopically distinguishable and thus nearly orthogonal states of the metric, each peaked around $g^{(i)}$ \cite{zych_2019,Christodoulou_2019, giacomini2021einsteins}. Formally, using a control qubit whose orthogonal states $\ket{1}$ and $\ket{2}$ distinguish the branches, the quantum state is
\begin{equation}
\alpha \ket{1}\ket{g^{(1)}}  + \beta \ket{2}\ket{g^{(2)}},\ \alpha,\beta\in\mathbb{C}, \label{eq: SP of metrics}
\end{equation} with $|\alpha|^2+|\beta|^2=1$ \footnote{Note that Eq.~\eqref{eq: SP of metrics} should be understood as suggestive since the Hilbert space assigned to the metric has not properly been introduced. However, if classically distinguishable spacetimes are represented by orthogonal states, one can form linear superpositions thereof and define a suitable inner product in the Hilbert space. They would correspond to the superposition of semi-classical states in a complete theory.}. Such a scenario might, for example, arise when placing a gravitating object in a superposition of two locations \cite{Marletto_2017, Bose_2017, Belenchia_2018, Christodoulou_2019, Anastopoulos_2020, zych_2019, delahamette2021falling, aspelmeyer2021}. Note that the states $\ket{g^{(i)}}$ in Eq.~\eqref{eq: SP of metrics} may be thought of as arising from semi-classical states defined at a spacelike surface peaked on initial data for the three--metric and extrinsic curvature such that classically through the equations of motion they generate the four--metrics $g^{(1)}$ and $g^{(2)}$ (cf.~\cite[Ch.~43]{Misner1973}). Moreover, in a covariant formulation, the configurations $g^{(i)}$ can be understood to arise from a stationary phase approximation in the path integral formalism \cite{Bengyat_2024}.

\begin{center}
\begin{figure*}
    \includegraphics[width=0.7\textwidth]{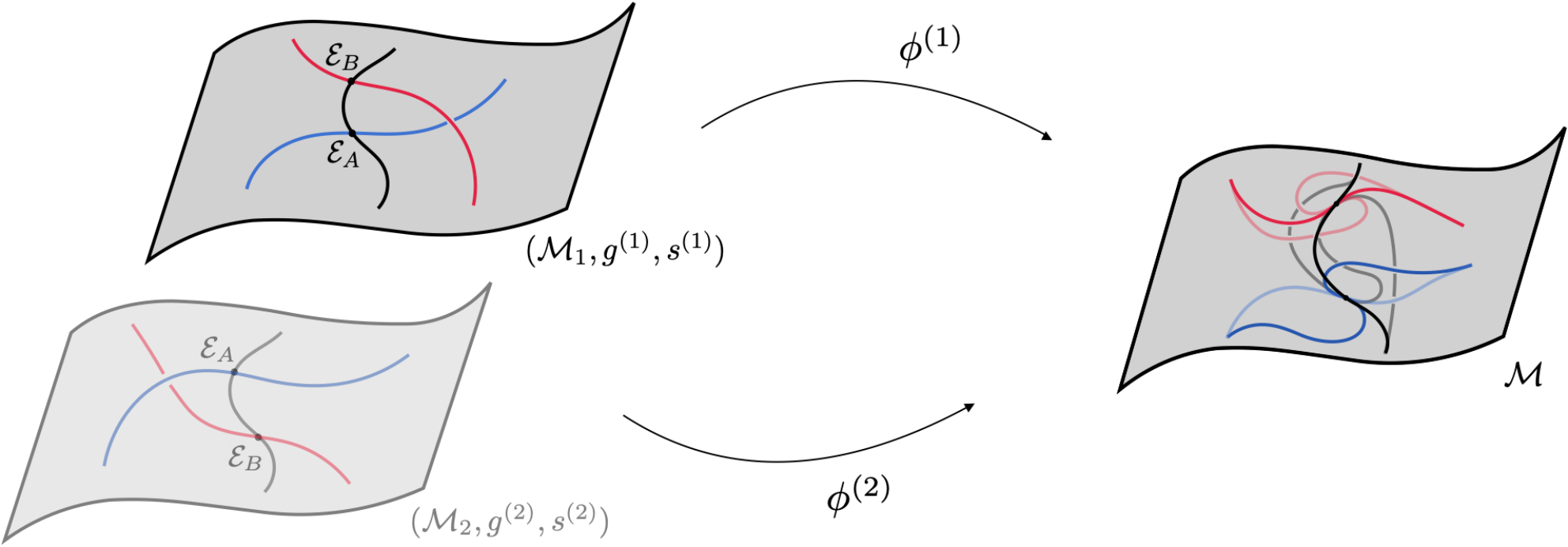}
    \caption{A superposition of two scenarios with different causal orders $s^{(1)}$ and $s^{(2)}$. In each branch, we consider a manifold $\mathcal{M}_{1,2}$, a metric $g^{(1),(2)}$, and the worldlines of three systems. The black, blue, and red line depict $\gamma_0$, $\gamma_A$, and $\gamma_B$, with different transparencies illustrating the superposition. There are \textit{two} events $\mathcal{E}_A$ and $\mathcal{E}_B$ defined by the crossing of the test particle with one of the other trajectories. First, we apply a quantum-controlled diffeomorphism $\phi^{(1)}: \mathcal{M}_1 \to \mathcal{M}$ together with a diffeomorphism $\phi^{(2)}: \mathcal{M}_2\to\mathcal{M}$ chosen such that each of the events is associated with a single point on $\mathcal{M}$. The metric and causal order may still be in superposition.}
    \label{fig:diffeo1}
\end{figure*}
\end{center}

Next, in order to characterize the causal order between two events on a superposition of spacetimes, we first need to define the notion of an event itself in this context. Since $\mathcal{E}_X^{(1)}$ and $\mathcal{E}_X^{(2)}$, $X=A,B$ both represent the crossing of the worldline of the test particle with that of system $X$, we posit that they represent the \emph{same} physical event:
\begin{align}
\mathcal{E}_X^{(1)}=\mathcal{E}_X^{(2)}\equiv\mathcal{E}_X, \ X=A,B.
\end{align}
This extends the relativistic notion of coincidences to incorporate superpositions of spacetimes. Note that it differs essentially from the notion of an event as a point on a spacetime manifold. This becomes particularly clear when counting the events involved in our scenario.
There are \emph{four} different \emph{manifold-points} involved in our scenario -- $p_{\mathcal{E}_A}^{(1)}$ and $p_{\mathcal{E}_B}^{(1)}$ on manifold $\mathcal{M}_1$ as well as $p_{\mathcal{E}_A}^{(2)}$ and $p_{\mathcal{E}_B}^{(2)}$ on manifold $\mathcal{M}_2$ (see Fig.~\ref{fig:diffeo1}). However, there are only \emph{two} distinct \emph{physical events} $\mathcal{E}_A$ and $\mathcal{E}_B$ -- the crossing of the test particle with system $A$ and the crossing of the test particle with system $B$.
\emph{Causal order on a superposition of spacetimes ---} We can now extend Definition \ref{def:CO} of causal order between events to superpositions of spacetimes.

\begin{definition}\label{def: ico}[Definite and indefinite causal order]
Consider a superposition of spacetimes $(\mathcal{M}_i,g^{(i)})$, $i\in\{1,2\}$ and two quantum events $\mathcal{E}_A$ and $\mathcal{E}_B$ connected by a timelike worldline $\gamma_0$. In each branch $i$ of the superposition, let $s^{(i)}$ denote the sign of the proper time difference between the crossing points $p_{\ea}^{(i)} = \gamma_0^{(i)}(\tau_A^{(i)})$ and $ p_{\eb}^{(i)} = \gamma_0^{(i)}(\tau_B^{(i)})$. If
\begin{equation}
s^{(1)} s^{(2)} = 1,
\end{equation}
we say that we have definite causal order between the two events $\ea$ and $\eb$. When
\begin{equation}
s^{(1)} s^{(2)} = -1,
\end{equation}
we say that we have indefinite causal order (ICO)\footnote{Note that this notion of causal order can be generalized straightforwardly to encompass the causal ordering of any finite number $N$ of timelike events in an arbitrary number of branches $M$. Denoting the sign of the proper time difference between two events $\mathcal{E}_l$ and $\mathcal{E}_k$ by $s_{lk}=\text{sign}(\tau_k-\tau_l)$, we say that a process has \emph{definite} causal order if \emph{all} pairwise products $s^{(i)}_{lk}s^{(j)}_{lk}, 1\leq l<k\leq N,1\leq i<j\leq M $ are positive. Otherwise, the process displays some form of ICO.}. 
\end{definition}

Thus, the notions of definite and indefinite causal order are conveniently encoded in the product $s^{(1)}s^{(2)}$. Importantly, ICO in the above sense can be due to a superposition of spacetime metrics $g_{\mu\nu}(x)$ or of crossing points $p_{\ea}$ and $p_{\eb}$\footnote{When considering ICO on a fixed spacetime background due to a superposition of crossing points, we consider two copies of the same manifold and metric, $(\mathcal{M}_1,g^{(1)}) = (\mathcal{M}_2, g^{(2)})$ in Definition 2.}. In the following, we will refer to the latter as ICO due to \emph{delocalized events} as, in this case, $\ea$ and $\eb$ are assigned different manifold points across the different branches. The above observation follows straightforwardly from Eq.~\eqref{eq:properTime}, which tells us that the proper time difference depends on both the metric along the curve $\gamma_0$ and the coincidences of worldlines that define the events $\ea$ and $\eb$. \emph{Thus, this notion of ICO captures both the optical and the gravitational quantum switch.} 

In the former, the causal order $s^{(i)}$ differs across the branches due to the superposition of paths of the target system. In the latter, this difference is due to the indefinite spacetime metric sourced by a superposed massive object. Using the above definition of ICO, \emph{there is no difference in the indefiniteness of the causal order between the optical and the gravitational quantum switch.}

This does not mean, of course, that there are no other observables that can distinguish between the two scenarios. Were one to measure the curvature scalar, for example, one would find that it is zero in the optical and non-zero in the gravitational quantum switch in the realization of \cite{zych_2019}. Nevertheless, having clarified what we mean by ICO in a general relativistic language, we find that both implementations lead to the same indefiniteness of causal order. In this sense, the optical quantum switch can be seen as just as much a \emph{realization} of ICO as a hypothetical realization of the gravitational quantum switch. In particular, this implies that quantum spacetime is not a necessary resource for a `genuine' realization of the quantum switch. 

\emph{Invariance under quantum diffeomorphisms ---} The quantity $s^{(1)}s^{(2)}$ is a meaningful observable in the general relativistic sense, that is, it is invariant under both classical and quantum diffeomorphisms. Since the proper time $\tau_{\gamma}$ is a diffeomorphism-invariant quantity, so is the sign of the proper time difference $s$ and this holds true in each branch of the superposition. Thus, the quantity $s^{(1)}s^{(2)}$ cannot be changed by a classical diffeomorphism. 

However, a classical diffeomorphism is not the most general symmetry transformation one can consider. Each of the classical configurations in Eq.~\eqref{eq: SP of metrics} can be described in a specific coordinate system. Considering superpositions thereof, we can choose a different coordinate system in each branch. These different choices of coordinates can be tied together with the help of physical events \cite{giacomini2021einsteins, giacomini2021quantum} to furnish a quantum coordinate system. Specifically, we use the term \emph{quantum coordinates} to refer to a pair of coordinates $(x^{(1)}_\mu$, $x^{(2)}_\mu)_{\mu=0,1,2,3}$, one for each branch $i\in\{1,2\}$ of the superposition. A coherent \emph{change} of quantum coordinates can then be implemented abstractly by a quantum-controlled pair of diffeomorphisms $\phi^{(1)}, \phi^{(2)}$ acting correspondingly on each branch -- a \emph{quantum diffeomorphism}.
Formally, this transformation can be written as
\begin{align}
    \ket{1}\bra{1}\otimes \mathcal{U}_1 + \ket{2}\bra{2}\otimes \mathcal{U}_2,
\end{align}
where $\mathcal{U}_{i}$ are unitary representations of the diffeomorphisms $\phi^{(i)}$.
Since the quantity $s^{(i)}$ is invariant under any classical diffeomorphism, it remains unchanged in each branch of the superposition, independently of the concrete diffeomorphism $\phi^{(i)}$ that is applied. Thus, the product $s^{(1)}s^{(2)}$ is invariant under any quantum diffeomorphism.

Note, however, that whether a process displays ICO `due' to delocalized events or a superposition of spacetime metrics can in general change under quantum diffeomorphisms. To see this, consider a scenario with ICO. Let us assume that we begin in a quantum coordinate system in which ICO is due to delocalized events while the spacetime background is fixed. More specifically, we assume that the crossing points $p_{\ea}^{(1)}, p_{\eb}^{(1)}$ and  $p_{\ea}^{(2)}, p_{\eb}^{(2)}$ differ across the branches of the superposition. Now, we can always find a quantum diffeomorphism that maps both configurations onto the same manifold $\mathcal{M}$ while ensuring that the events $\ea$ and $\eb$ are each associated with a single point on this manifold (see also \cite{Oreshkov_2012}). That is, there always exist diffeomorphisms $\phi^{(1)}: \M_1 \to \M$ and $\phi^{(2)}: \M_2 \to \M$ such that 
\begin{align}
    \phi^{(1)}(p_{\ea}^{(1)}) &= \phi^{(2)}(p_{\ea}^{(2)}) \equiv p_{\ea},\\
    \phi^{(1)}(p_{\eb}^{(1)}) &= \phi^{(2)}(p_{\eb}^{(2)}) \equiv p_{\eb}.
\end{align}
Thus, whether the events $\ea$ and $\eb$ are localized or delocalized depends on the choice of quantum coordinates. If we had defined events in terms of spacetime points (as suggested e.g.~in \cite{Paunkovic_2020}), their number would change under a quantum diffeomorphism. But, as we have seen, the causal order itself is invariant under quantum diffeomorphisms; therefore, the setup must still exhibit ICO. This is because a quantum diffeomorphism also affects the metric. If we start with a definite spacetime metric $g$ and apply a different diffeomorphism in each branch, we end up with a superposition of spacetime metrics $g^{(i)} = (\phi^{(i)})^{-1}_\ast(g)$, $i\in\{1,2\}$\footnote{Note that the action of the diffeomorphisms on the fields is the inverse of their actions on the manifold points.}. \emph{In this sense, a quantum diffeomorphism can shift the indefiniteness due to delocalized events into an indefiniteness of the spacetime metric at localized events.} A concrete example of this transformation is provided in Supplemental Material \ref{app:delevents}.

Conversely, if we start with a superposition of different spacetime metrics $g^{(1)}$ and $g^{(2)}$ and assume that the test particle moves along a geodesic in each branch, we can always find a quantum diffeomorphism that renders the metric flat and thereby definite along the entire path of the test particle. This is because, in each branch of the superposition, there exist Fermi normal coordinates such that the metric is locally flat along the geodesic (\cite[Sec.~13.6]{Misner1973},\cite{giacomini2021einsteins,giacomini2021quantum}). As a result, the metric becomes definite in the region relevant to the computation of the proper time $\tau_{\gamma_0}$ (Eq.~\eqref{eq:properTime}) and the indefiniteness of causal order is now due to a superposition of worldline crossings. A concrete construction can be found in Supplemental Material \ref{app:supSP}. In short, whether ICO is due to delocalized events or a superposition of metrics -- a key difference between the optical and gravitational implementation of the quantum switch -- \emph{depends on the choice of quantum coordinates.} Therefore, in the context of the quantum switch, there is no absolute matter of fact whether ICO is due to quantum features of spacetime.

Moreover, note that at least one of the two -- the spacetime metric or the location of events -- must be in a quantum superposition to verify ICO. There is no way to generate ICO, as defined in Definition \ref{def: ico}, in a definite spacetime background with localized events. We can straightforwardly phrase this in terms of a no-go theorem. There is no quantum coordinate system in which the following three statements hold:
\begin{enumerate}
    \item Both events $\ea$ and $\eb$ are localized.
    \item The spacetime metric is definite.
    \item The causal order between $\ea$ and $\eb$ is indefinite.
\end{enumerate}
This is consistent with the results of \cite[Corollary 1]{Vilasini_2022}, despite the different formalisms for ICO.

\emph{Operational encoding as quantum mechanical observable ---} The quantity $s$ characterizing causal order is also an observable in the quantum mechanical sense. To show this, we provide a concrete operational encoding of $s$ in orthogonal qubit states $\ket{s=\pm 1}$ for a setup displaying a superposition of causal orders (see Fig.~1 in Supplemental Material \ref{app:encoding}). Assume that the test particle has an internal spin degree of freedom, which precesses around the $z$-axis according to its proper time \cite{Zych_2011, Smith_2020natcomm, Paczos_2022, Debski_2022} and is prepared in the initial state $\ket{b_0}$ in both branches. In addition, consider the systems $A$ and $B$ to be 
Alice's and Bob's laboratories respectively, with $\gamma_A$ and $\gamma_B$ their worldlines. We want to ensure that the particle crosses the laboratories at specific proper times. Its first crossing with a laboratory occurs at $\tau^*_1$, the second crossing at  $\tau^*_2$. Thus, in the first branch, $p_{\mathcal{E}_A}^{(1)}=\gamma_0(\tau^*_1)$ and $p_{\mathcal{E}_B}^{(1)}=\gamma_0(\tau^*_2)$, while in the second branch, $p_{\mathcal{E}_B}^{(2)}=\gamma_0(\tau^*_1)$ and $p_{\mathcal{E}_A}^{(2)}=\gamma_0(\tau^*_2)$.

When the spin degree of freedom enters the first laboratory at proper time $\tau^*_1$, it has evolved to be in some state $\ket{b_1}$. The crossings of the worldlines are now tuned such that the spin is in the orthogonal state $\ket{b_2}$ when it crosses the second laboratory at time $\tau^*_2$. This ensures that, whenever the particle enters a laboratory, the corresponding agent can measure in the basis $\{ \ket{b_1}, \ket{b_2} \}$ without disturbing the state and its time evolution. Upon measurement, each agent encodes either $\tau_1^\ast$ or $\tau_2^\ast$, depending on the result of their measurement $b_1$ or $b_2$ respectively, in a memory register with associated  Hilbert space $\mathcal{H}_{A,B}\cong \mathbb{C}^2$, initialized to $\ket{0}_{A,B}$. Thus, the overall state evolves from the initial state
\begin{align}
    \ket{\psi_1}=(\alpha\ket{1}\ket{g^{(1)}}+\beta\ket{2}\ket{g^{(2)}})\ket{b_0}\ket{0}_A\ket{0}_B\nonumber
\end{align}
to the state after the particle enters the first laboratory,
\begin{align}
    \ket{\psi_2}=(\alpha\ket{1}\ket{g^{(1)}}\ket{\tau^*_1}_A\ket{0}_B+\beta\ket{2}\ket{g^{(2)}}\ket{0}_A\ket{\tau^*_1}_B)\ket{b_1},\nonumber
\end{align}
and finally, after it has entered the second laboratory, to
\begin{align}
    \ket{\psi_3}=(\alpha\ket{1}\ket{g^{(1)}}\ket{\tau^*_1}_A\ket{\tau^*_2}_B+\beta\ket{2}\ket{g^{(2)}}\ket{\tau^*_2}_A\ket{\tau^*_1}_B)\ket{b_2}.\nonumber
\end{align}
During post-processing, once the spin state has evolved to $\ket{b_f}$, a referee combines the two memory states and unitarily transforms the overall state to
\begin{align}
(\alpha\ket{1}&\ket{g^{(1)}}\ket{\tau^*_2-\tau^*_1}_A\ket{\tau^*_1+\tau^*_2}_B+\nonumber\\\beta\ket{2}&\ket{g^{(2)}}\ket{\tau^*_1-\tau^*_2}_A\ket{\tau^*_2+\tau^*_1}_B)\ket{b_f}.
\end{align}
Now, the causal order $s$ can be encoded in a quantum state by defining $\ket{s=\pm 1}:=\ket{\pm (\tau^*_2-\tau^*_1)}_A$. Since the state of register $B$ and the spin degree of freedom factorize out, they can be traced out:
    \begin{align}
        \ket{\Psi} = \alpha\ket{1} \ket{g^{(1)}}\ket{s=+1} +\beta \ket{2} \ket{g^{(2)}}\ket{s=-1}.
    \end{align}
Now, $s$ can be determined by measuring the control, together with the spacetime metric, in the basis $\{\ket{\phi_{\pm}}=\frac{1}{\sqrt{2}}(\ket{1}\ket{g^{(1)}} \pm \ket{2}\ket{g^{(2)}})\}$. By post-selecting on the outcome $\ket{\phi_+}$, the test particle is left in a superposition of states corresponding to opposite causal orders,
\begin{align}
    \ket{\psi} = \alpha\ket{s=+1} +\beta \ket{s=-1}.
    \label{eq:ICOstate}
\end{align}

The same procedure can be applied in the case of a definite causal order. It is easy to see that in these cases, the final state before post-selection becomes $(\alpha\ket{1}\ket{g^{(1)}} +\beta\ket{2}\ket{g^{(2)}})\ket{b_f} |s=\pm 1\rangle$. In Supplemental Material \ref{app:encoding}, we provide details on a measurement scheme to determine the state of the test particle and distinguish between definite and coherently or incoherently superposed causal order.

\emph{Discussion ---} We proposed a notion of ICO between operationally defined events. One key conceptual contribution is the extension of the notion of an event to superpositions of spacetimes. The advantage of this notion is that it is independent of the choice of classical and quantum coordinates and thus qualifies as an observable in the general relativistic sense. We further showed that the proposed notion of ICO constitutes an observable in the quantum mechanical sense, and that this observable describes both the optical and the gravitational quantum switch. In particular, we have seen that whether causal order is definite or indefinite is invariant under quantum coordinate transformations. On the other hand, whether ICO is due to delocalized events or a superposition of spacetimes is a matter of choice of quantum coordinates. Thus, we argue that the optical quantum switch can be seen as a \emph{realization} of indefinite causal order rather than a simulation. Note that this implies that a realization of ICO does not suffice to show that spacetime has been set in a superposition. Moreover, we have seen that there is no quantum coordinate system in which (i) both events $\ea$ and $\eb$ are localized, (ii) the spacetime metric is definite, and (iii) the causal order between these events is indefinite (cf.~\cite{Vilasini_2022}).

The present work connects scenarios with localized events on an indefinite spacetime background with delocalized events on definite spacetime. This idea has been prevalent in the community for years \cite{Oreshkov_2019, castroruiz2019time,giacomini2021einsteins, giacomini2021quantum, delahamette2021falling, Kabel2022conformal, Vilasini_2022, Paunkovic_2020} but is made explicit here. We are positive that emphasizing this connection will facilitate the discussion between the general relativity and quantum information communities and shed light on the interplay between indefinite causality and the quantum nature of spacetime. This might also be relevant to the ongoing research efforts regarding potential tests of quantum features of gravity in the low energy regime (e.g.~\cite{Marletto_2017, Bose_2017, Belenchia_2018, Christodoulou_2019, Anastopoulos_2020, zych_2019, delahamette2021falling, aspelmeyer2021}).

What remains to be done is to cast the quantum-controlled diffeomorphisms employed here in the framework of the quantum reference frame formalism. This requires identifying the explicit form of the unitary representations $\mathcal{U}_{1,2}$ and the corresponding Hilbert space structure for the spacetime metric and the quantum coordinates. A promising route to constructing such explicit QRF transformations currently explored is to extend the frameworks of \cite{giacomini2021einsteins, giacomini2021quantum, delahamette2021falling}. Another interesting direction for future research is to extend the notion of ICO to more general scenarios. This includes finding an appropriate definition of event when given worldlines with multiple crossings or even different numbers of crossings across the branches of the superposition, as well as generalizing the notion to quantum gravitational scenarios beyond a superposition of semi-classical spacetimes. In particular, an important open question is concerned with the extension of the notion of event to genuine quantum fields. In this context, it would further be interesting to explore the relation between ICO and micro-causality in algebraic quantum field theory (cf.~\cite{Ormrod2023causal}, \cite[Ch.~3]{Weinberg_1995}, \cite[Ch.~3]{StreaterWightman_2001}).

The present findings are in line with several other works on ICO. First, we can view them in light of recent results in the process matrix formalism. It was shown in \cite{Castro-Ruiz17} that the causal order between operations is always preserved under continuous and reversible transformations (see also \cite{Selby_2020,castroruiz2020comment} for further discussion). This result is derived in an abstract framework of Hilbert spaces in the absence of any spacetime structure and has not yet been extended to a field theoretic realm. However, similarities between the general structure of the transformations in this work and diffeomorphisms hint at a deeper connection, which could explain why causal order is preserved in both frameworks.
Second, different notions of ICO have also been analyzed and compared in \cite{Vilasini_2022}. The authors connect information-theoretic notions of ICO to relativistic causality and prove several no-go results in the context of localized events, definite spacetime, and classical reference frames. We have already seen consistency with one of their results in the aforementioned no-go-theorem. It would be interesting to investigate to what extent these results extend further to the realm considered here of delocalized events, indefinite metrics, and quantum coordinate systems. \\

\begin{acknowledgments}
MC acknowledges several useful preliminary discussions with Giulio Chiribella and Some Sankar Bhattacharya. VK acknowledges support through a DOC Fellowship of the Austrian Academy of Sciences. This research was funded in whole or in part by the Austrian Science Fund (FWF) [10.55776/F71]. This publication was made possible through the support of the ID 61466 and ID 62312 grants from the John Templeton Foundation, as part of The Quantum Information Structure of Spacetime (QISS) Project (qiss.fr). The opinions expressed in this publication are those of the authors and do not necessarily reflect the views of the John Templeton Foundation.\\
\end{acknowledgments}

\nocite{apsrev42Control}
\bibliography{bibliography}
\bibliographystyle{apsrev4-2.bst}

\appendix
\onecolumngrid
\section*{Supplemental Material}

\appendix\label{sec: Supplementary Note}

\section{Mapping between different descriptions of the quantum switch}

In this Supplemental Material, we take a closer look at the concrete mappings from a description of the quantum switch in which ICO is due to delocalized events to one in which ICO is due to a superposition of spacetime metrics and vice versa.

\subsection{Delocalized events to superposition of spacetime metrics} \label{app:delevents}

To illustrate the first of these maps, let us start with a situation with indefinite causal order due to delocalized events on a fixed Minkowski background. Denoting the points associated to the events $\ea$ and $\eb$ in the respective branches of the superposition by $p_{\ea,\eb}^{(1)} \in \M_1$ and $p_{\ea,\eb}^{(2)} \in \M_2$, one can always find diffeomorphisms $\phi^{(1)}: \M_1 \to \M$ and $\phi^{(2)}: \M_2 \to \M$ such that 
\begin{align}
    \phi^{(1)}(p_{\ea}^{(1)}) &= \phi^{(2)}(p_{\ea}^{(2)}) \equiv p_{\ea},\\
    \phi^{(1)}(p_{\eb}^{(1)}) &= \phi^{(2)}(p_{\eb}^{(2)}) \equiv p_{\eb},
\end{align}
while smoothly connecting to the identity outside of a compact region containing these points.
In this way, the events are \emph{localized}, that is, they are assigned the same respective point on the manifold $\mathcal{M}$, while the initial and final points $p_{\gamma_0,i}$ and $p_{\gamma_0,f}$ of the worldline $\gamma_0$ remain localized. Note that the idea that one can find coordinates such that the events $\ea$ and $\eb$ -- in this case defined by the application of operations in $A$ and $B$'s laboratory -- become localized was also discussed in \cite{Oreshkov_2016_Cerf}. Let us also briefly comment on a common misunderstanding: one might worry that no diffeomorphism can map two different points onto the same one since diffeomorphisms are one-to-one maps. This reasoning does not apply to the present case since we are taking points $p_{\ea,\eb}^{(1)} \in \M_1$ and $p_{\ea,\eb}^{(2)} \in \M_2$ from two \emph{different} manifolds $\M_1$ and $\M_2$ onto the same point $p_{\ea,\eb} \in \mathcal{M}$. 

Importantly, these diffeomorphisms $\phi^{(1)}$ and $\phi^{(2)}$ also act on the spacetime metric, originally described by the Minkowski metric $\eta$ across both branches, as
\begin{align}
    \big((\phi^{(1)})^{-1}_\ast(\eta)\big)(p_\ea) &\equiv g^{(1)}(p_\ea),\ \big((\phi^{(2)})^{-1}_\ast(\eta)\big)(p_\ea) \equiv g^{(2)}(p_\ea),\\
    \big((\phi^{(1)})^{-1}_\ast(\eta)\big)(p_\eb) &\equiv g^{(1)}(p_\eb),\ \big((\phi^{(2)})^{-1}_\ast(\eta)\big)(p_\eb) \equiv g^{(2)}(p_\eb),
\end{align}
leading to a superposition of metric field values at the points $p_\ea$ and $p_\eb$, respectively. Note that the metrics $g^{(1)}$ and $g^{(2)}$ are diffeomorphic, namely, they are related by $g^{(2)}={(\phi^{(1)} \circ (\phi^{(2)})^{-1})}_\ast (g^{(1)})$. Nevertheless, because the metric field \emph{at a particular spacetime point} is not a diffeomorphism-invariant quantity, it ends up in a superposition of values at this particular point. In fact, the metric field can even be in a superposition of different values along the entire worldline $\gamma_0$ between the two events. Diffeomorphism-invariant global geometric properties, on the other hand, are unaffected by this transformation. The two metrics in superposition still both describe a flat spacetime. However, the \emph{overall} configurations, including both the metric and the worldlines, across the two branches of the superposition are \emph{not} related by a diffeomorphism. In particular, the diffeomorphism-invariant quantity $s$, that is, the causal order in each branch, is different, since we \emph{started} with the causal order in a superposition. This fact remains unchanged. If we had started with a situation with definite causal order in the original setup, then no transformation could have made ICO emerge.

\subsection{Superposition of spacetime metrics to delocalized events}\label{app:supSP}

To illustrate the second map, let us start with a situation with indefinite causal order due to a superposition of spacetime metrics $g^{(1)}$ and $g^{(2)}$ associated with $\M_1$ and $\M_2$, respectively, while the events are localized. In particular, we may consider two metric fields that are not related to one another by a diffeomorphism. Nevertheless, \emph{along} the worldline of the test particle, we can choose Fermi normal coordinates in each branch of the superposition. These coordinates are defined such that the metric is locally flat along the entire worldline (\cite[Sec.~13.6]{Misner1973},\cite{giacomini2021einsteins,giacomini2021quantum}). Due to the close connection between coordinates and diffeomorphisms, there exist diffeomorphisms $\phi^{(1)}: \M_1 \to \M$ and $\phi^{(2)}: \M_2 \to \M$ such that 
\begin{align}
    (\phi^{(1)})^{-1}_\ast(g^{(1)})|_{\gamma_0} =(\phi^{(2)})^{-1}_\ast(g^{(2)})|_{\gamma_0} = \eta|_{\gamma_0}.
\end{align}
These diffeomorphisms also act on the particle's worldline. In particular, they map $p_\ea$ and $p_\eb$ to two different points each:
\begin{align}
    \phi^{(1)}(p_\ea) &= p_\ea^{(1)}, \phi^{(2)}(p_\ea) = p_\ea^{(2)},\\
    \phi^{(1)}(p_\eb) &= p_\eb^{(1)}, \phi^{(2)}(p_\eb) = p_\eb^{(2)}.
\end{align}
Thus, the events $\ea$ and $\eb$ are \emph{delocalized}. Again, because it is a diffeomorphism-invariant quantity, the causal order remains indefinite. We have thus mapped a situation with ICO due to a superposition of spacetime metrics to a situation with ICO due to delocalized events. Note, importantly, that we have only rendered the metric flat and definite \emph{along the worldline $\gamma_0$}. While this is the only domain relevant for causal order defined through differences of proper time, $\tau_{\gamma_0} = \frac{1}{c}\int_{p_{\gamma_0,i}}^{p_{\gamma_0,f}}\sqrt{-g_{\mu\nu}dx^\mu dx^\nu}$, the metric generally remains in superposition of values at any point that does not belong to $\gamma_0$. Global and diffeomorphism-invariant geometric quantities will not be affected by this transformation.\\

Remark: In light of more recent work by some of us \cite{Kabel_2024}, a more explicit way of modeling the maps between manifolds in superposition would be to retain the distinction between manifolds $\M_1$ and $\M_2$ even after the application of the diffeomorphism and instead introduce a \enquote{comparison map} $C_\chi: \M_1 \to \M_2$, constructed from physical fields $\chi$ on the manifolds, that allows us to relate points in one branch of the superposition to points in the other branch and thus define the localization of events. The present results fit into this more general framework by acknowledging that we are implicitly using the identity to compare across the two branches of the superposition (see also \cite[p.~22]{Kabel_2024}).

\section{An operational encoding of the causal order}\label{app:encoding}

In this Supplemental Note, we discuss in further detail the operational encoding and verification of indefinite causal order. The protocol proposed in the section \enquote{\emph{Operational encoding as quantum mechanical observable}} of the main text is illustrated in Fig.~\ref{fig: ICOappendix} below.

\begin{figure}[h!]
    \centering
    \includegraphics[scale=0.34]{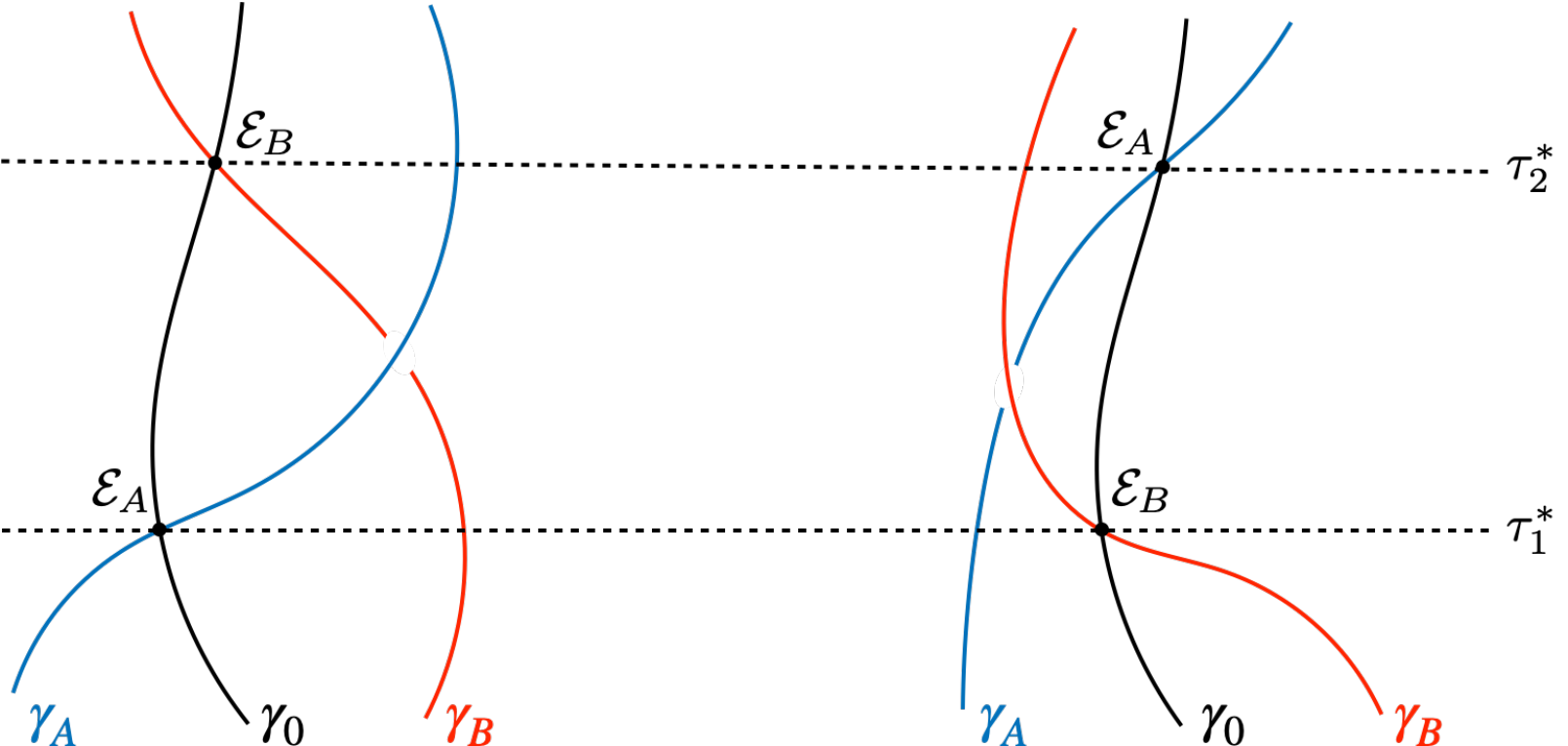}
    \caption{In order to encode the causal order operationally, we consider a particle with an internal spin degree of freedom moving along $\gamma_0$, and two agents in laboratories with worldlines $\gamma_A$ and $\gamma_B$. These are placed in a superposition of causal orders between events $\mathcal{E}_{A}$ and $\mathcal{E}_B$. The setup is chosen such that the particle always enters the first laboratory at proper time $\tau_1^\ast$ while it crosses the second laboratory at $\tau_2^\ast$. Through a careful choice of these proper times, we ensure that the agents can perform non-disturbing measurements on the particle whenever it enters their laboratory and thus encode the causal order in a memory register.}
    \label{fig: ICOappendix}
\end{figure}

\begin{figure}[h!]
    \centering
    \includegraphics[width=0.3\textwidth]{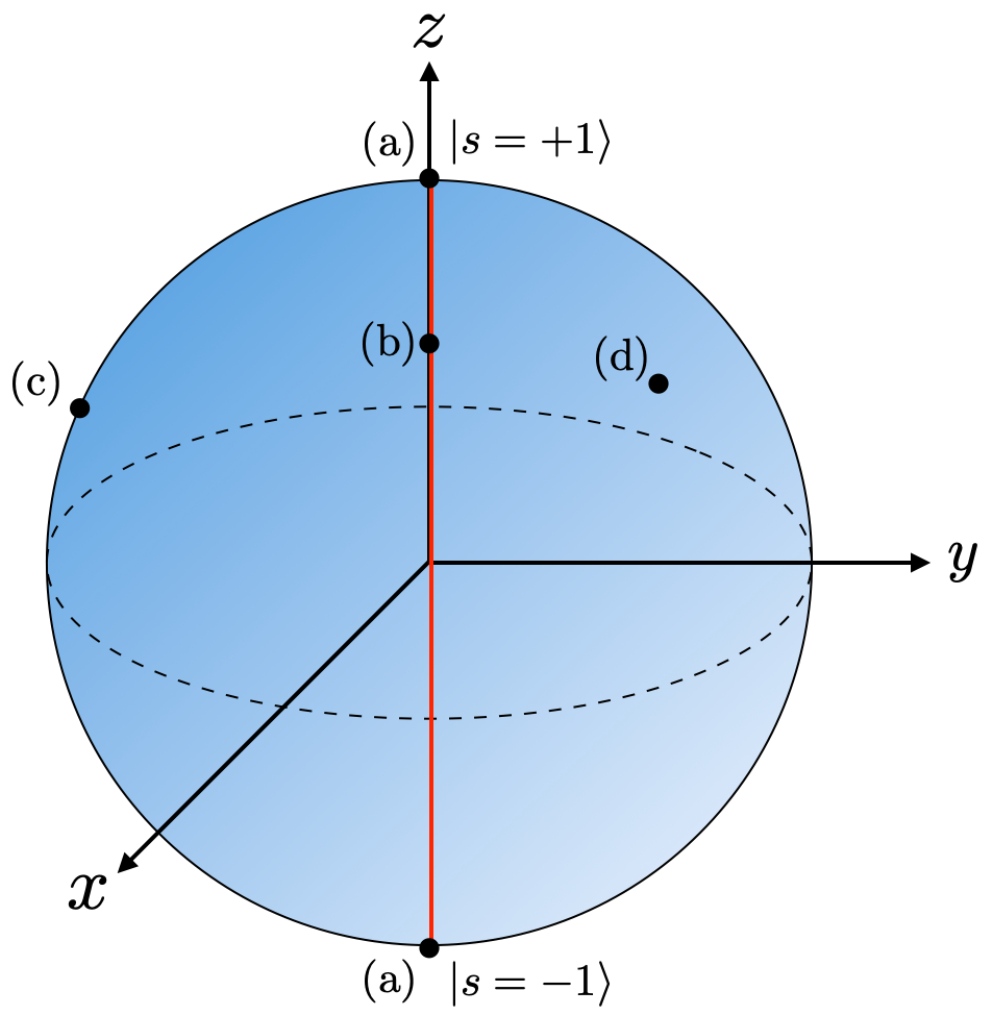}
    \caption{Bloch representation of the state space of operationally encoded causal order. Along the $z$-axis, the causal order is (a) definite, that is, in a state $\ket{s=\pm 1}$, or (b) in a classical mixture of such states. (c) On the surface of the Bloch ball (except the points $\ket{s=\pm 1}$),
    the system is in a coherent superposition of causal orders and, finally, (d) the remaining states inside the ball represent
    mixed indefinite causal orders.
    }
    \label{fig: bloch sphere order}
\end{figure}

\noindent As seen above, this protocol gives rise to the following state of the test particle:
\begin{align}
    \ket{\psi} = \alpha\ket{s=+1} +\beta \ket{s=-1},\ \alpha, \beta \in \mathbb{C}.
    \label{eq:ICOstate}
\end{align}
Collecting the statistical data of the measurements of $\sigma_z$, $\sigma_x$, and $\sigma_y$ on the test particle, one can distinguish between pure indefinite causal order as in Eq.~\eqref{eq:ICOstate} and definite causal orders, a classical mixture of definite (opposite) causal orders, and mixed indefinite causal orders (see Fig.~\ref{fig: bloch sphere order}). Note that the measurement of only $\sigma_z$ would not be sufficient because such a measurement would not distinguish coherently superposed causal orders from a classical mixture. Many other specific proposals for protocols that verify ICO can be found in the literature  \cite{Araujo_2015, Branciard_2015, Branciard_2016, Bavaresco_2019, vanderLugt_2022}. Going beyond the protocol given here, Alice and Bob may apply more general operations than the ones described here and thereby also signal in a superposition of orders.

\end{document}